\documentclass[12pt]{iopart}

\newcommand{\be}{\begin{eqnarray}}
\newcommand{\ee}{\end{eqnarray}}
\newcommand{\bea}{\begin{eqnarray}}
\newcommand{\eea}{\end{eqnarray}}

\usepackage{graphicx}
\usepackage{bm}
\usepackage{color}
\usepackage{slashed}
\usepackage{cite}
\usepackage{graphicx}
\usepackage{multicol}
\usepackage{graphicx}
\usepackage{color}
\usepackage{slashed}
\usepackage{CJKutf8}

\begin{document}
\begin{CJK}{UTF8}{<font>}
\title{Comparison of thermodynamic behaviors of two regular-AdS black holes}

\author{Sen Guo$^{1}$, \ Ya-Ling Huang$^{2}$, \ En-Wei Liang$^{1*}$}

\address{
$^1$Guangxi Key Laboratory for Relativistic Astrophysics, School of Physical Science and Technology, Guangxi University, Nanning 530004, People's Republic of China\\
$^2$School of Electrical Engineering, SouthWest JiaoTong University, Chengdu 610031, People's Republic of China}

\ead{sguophys@126.com; katrina996@163.com; lew@gxu.edu.cn}
\vspace{10pt}
\begin{indented}
\item[]Sep 2021
\end{indented}

\begin{abstract}
Considering the negative cosmological constant of an anti-de Sitter (AdS) background as a positive thermodynamic pressure in the extended phase space, we study the thermodynamic behaviors of the regular Hayward-AdS (HAdS) black hole (BH) by investigating $P-\upsilon$ critical and cooling-heating phase transition. We compare the difference of two thermodynamic processes between the HAdS BH and the Bardeen-AdS (BAdS) BH. It is found that the phase transition of the BAdS BH tends to be more the van der Waals (vdW) phase transition. For the cooling-heating phase transition, we find the inversion curves of the HAdS BH are always higher than the BAdS BH under the same pressure and magnetic charge. We also compare the smallest existence mass, the zero-temperature remnant, and the critical magnetic charge. The results suggest that the inner horizon and the outer horizon of the Hayward BH are easier to merge, and the singularity is easier to expose.
\end{abstract}


\section{Introduction}
\label{intro}
\par
Black hole (BH) singularity is a well-known challenge in general relativity. Penrose and Hawking's famous work showed that the existence of singularity is inevitable \cite{1}. In order to avoid singularity, the regular solution of the Einstein equation is constructed. Bardeen obtained the first regular BH solution in 1968 \cite{2}. Ay\'{o}n-Beato and Garc\'{\i}a pointed out the physical source of the regular BHs might be the nonlinear electrodynamics \cite{3}. Inspired by Bardeen's idea, Hayward proposed a static spherically symmetric BH sulution, showing the limitation and regularity invariance of the curvature in the regular space-time \cite{4}. The stress-energy tensor of the Hayward BH satisfies the weak energy condition, and it violates the strong energy condition and belongs to a degenerate configuration of the nonlinear magnetic monopole gravity field \cite{5}. Moreover, many kinds of researches showed that the regular BHs always share similar kinetic properties \cite{6,7,8,9,10,11}.

\par
Since Hawking radiation was first proposed in 1974, the thermodynamics properties of the BHs were extensively studied \cite{12,13,14}. The BH as a thermodynamic system has many exciting similarities with the classical thermodynamics system. These similarities become more evident and precise at the anti-de Sitter (AdS) space-time \cite{15}. Furthering studies showed that the phase transition of the charged AdS BH likes the van der Waals (vdW) liquids system \cite{16}. Significantly, an exciting proposal is to treat the negative cosmological constant of the AdS space-time background as a positive thermodynamic pressure in the extended phase space, i.e., $P\equiv -\Lambda/{8\pi}$, which can enrich the AdS BH thermodynamic researches \cite{17}. Based on this proposal, the thermodynamics properties of the AdS BHs were multifacetedly investigated, including $P-\upsilon$ critical behavior \cite{18,19,20,21,22,23,24}, Joule-Thomson expansion (JT) \cite{25,26,27,28,29}, weak cosmic censorship conjecture \cite{30,31,32}, and BH heat engine \cite{33,34}. By investigating the JT expansion of the BAdS BH, we analyzed the inversion features of this BH \cite{28}. By constructing the HAdS BH heat engine, we discussed the relationship between the HAdS BH and the BAdS BH. \cite{35}. Furthermore, the BAdS BH phase transition grade from both macroscopic and microscopic was studied, we found that the BAdS BH phase transition should be a second-order phase transition \cite{L}.

\par
Based on these analyses, we naturally ask whether the HAdS BH and the BAdS BH exhibit different thermodynamic properties from the singular BH. For these two regular BHs, their respective thermodynamic properties have also aroused our interest. We study these issues in this paper. We investigate the $P-\upsilon$ critical and JT expansion of the HAdS BH and compare the differences between the HAdS BH and the BAdS BH for these two thermodynamic processes. Furthermore, we calculate the smallest existence mass, the zero-temperature remnants, and the critical magnetic charge of the HAdS and the BAdS BHs. The outlines of this paper are listed as follows. In Sec.2, we briefly review the regular HAdS and BAdS BHs thermodynamic. Sec.3 investigates the $P-\upsilon$ critical and the cooling-heating phase transition of these two regular BHs. Sec.4, we analyze the thermodynamic differences between the HAdS and BAdS BHs from three perspectives. Sec.5 ends up with conclusions and discussions. For simplicity, we adopt the dimensionalization as $G_{\rm N}=\hbar=\kappa_{\rm B}=c=1$.

\section{Thermodynamic quantities and the equation of the state}
\label{sec2}
\par
The regular-AdS BH metric is given by \cite{36}
\begin{equation}
\label{2-1}
{\rm d}s^{2}=-f(r){\rm d}t^{2}+\frac{{\rm d}r^{2}}{f(r)}+r^{2}{\rm d}\Omega^{2}.
\end{equation}
The HAdS and BAdS BHs metric potential can be expressed by \cite{37,38,39}
\begin{equation}
\label{2-2}
f_{\rm H}(r)=1+\frac{r^{2}}{l^{2}}-\frac{2 M r^{2}}{r^{3}+g^{3}},
\end{equation}
\begin{equation}
\label{2-3}
f_{\rm B}(r)=1+\frac{r^{2}}{l^{2}}-\frac{2 M r^{2}}{(r^{2}+g^{2})^{3/2}},
\end{equation}
where $M$ is BH mass, $g$ is the magnetic charge of the BH, and $l$ is AdS radius. `$\rm H$' and `$\rm B$' in the subscript stands for the HAdS BH and the BAdS BH, respectively. Allowing us to define conserved mass-energy for the event horizon ($r_{\rm +}$) without taking into account the inner horizon ($r_{\rm -}$). The first law of thermodynamics of the BH can be written as
\begin{equation}
\label{aa}
{\rm d}M=T{\rm d}S+V {\rm d}P+\varphi {\rm d}g.
\end{equation}
According to the first law of thermodynamics and $P\equiv -\Lambda/{8\pi}=3/8\pi l^2$, the thermodynamic variables of the HAdS BH and BAdS BH are obtained, i.e.,

\textit{\rm HAdS BH}
\begin{equation}
\label{2-4}
M_{\rm H}=\frac{(l^{2}+r_{\rm +}^{2})(g^3+r_{\rm +}^3)}{2l^{2}r_{\rm +}^{2}},~~~V_{\rm H}=\frac{4}{3}\pi (g^{3}+r_{\rm +}^{3}),
\end{equation}
\begin{equation}
\label{2-5}
T_{\rm H}=\frac{3r_{\rm +}^{5}+l^{2}(r_{\rm +}^{3}-2g^{3})}{4 \pi l^{2} r_{\rm +}(g^3+r_{\rm +}^{3})},~~~\varphi_{\rm H}=\frac{3g^{2}(l^{2}+r_{\rm +}^{2})}{2l^{2}r_{\rm +}^{2}}.
\end{equation}

\textit{\rm BAdS BH}
\begin{equation}
\label{2-6}
M_{\rm B}=\frac{(l^{2}+r_{\rm +}^{2})(g^2+r_{\rm +}^2)^{3/2}}{2l^{2}r_{\rm +}^{2}},~~~V_{\rm B}=\frac{4}{3}\pi (g^{2}+r_{\rm +}^{2})^{3/2},
\end{equation}
\begin{equation}
\label{2-7}
T_{\rm B}=\frac{3r_{\rm +}^{4}+l^{2}(r_{\rm +}^{2}-2g^{2})}{4 \pi l^{2} r_{\rm +}(g^2+r_{\rm +}^{2})},~~~\varphi_{\rm B}=\frac{3g(l^{2}+r_{\rm +}^{2})\sqrt{g^2+r_{\rm +}^{2}}}{2l^{2}r_{\rm +}^{2}}.
\end{equation}
Using above equations (\ref{2-4})-(\ref{2-7}), the equation of the state of the HAdS BH and the BAdS BH can be written as
\begin{equation}
\label{2-8}
P_{\rm H}=\frac{T}{2 r_{\rm +}}-\frac{1}{8 \pi {r_{\rm +}}^{2}}+\frac{g^3 T}{2r_{\rm +}^{4}}+\frac{g^{3}}{4 \pi {r_{\rm +}}^{5}},
\end{equation}
\begin{equation}
\label{2-9}
P_{\rm B}=\frac{T}{2 r_{\rm +}}-\frac{1}{8 \pi {r_{\rm +}}^{3}}+\frac{g^2 T}{2r_{\rm +}^{3}}+\frac{g^{2}}{4 \pi {r_{\rm +}}^{4}}.
\end{equation}

\par
Moreover, these two regular BHs' heat capacities at constant pressure/volume and the Gibbs free energy are obtained according to equations (\ref{2-4})-(\ref{2-7}), respectively. The heat capacity at constant volume as
\begin{equation}
\label{2-10}
C_{\rm HV}=C_{\rm BV}=T\Big(\frac{{\rm d} S}{{\rm d} T}\Big)_{\rm V,g}=0.
\end{equation}
The heat capacity at constant pressure can be written as
\begin{equation}
\label{2-11}
C_{\rm HP}=T\Big(\frac{{\rm d} S}{{\rm d} T}\Big)_{\rm P,g}=\frac{2 \pi r_{\rm +}^{2}(3r_{\rm +}^{5}-2 l^{2}g^{3}+r_{\rm +}^{3}l^{2})}{3r_{\rm +}^{5}+8l^{2}g^{3}-l^{2}r_{\rm +}^{3}},
\end{equation}
and
\begin{equation}
\label{2-12}
C_{\rm BP}=T\Big(\frac{{\rm d}S}{{\rm d}T}\Big)_{\rm P,g}=\frac{2 \pi r_{\rm +}^{2}(g^{2}+r_{\rm +}^{2})(3r_{\rm +}^{4}-2l^{2}g^{2}+l^{2}r_{\rm +}^{2})}{3r_{+}^{6}+l^{2}(8g^{4}+4g^{2}r_{\rm +}^{2}-r_{\rm +}^{4})}.
\end{equation}
In the extended phase space, the BH mass can be regarded as enthalpy $H$, hence, the Gibbs free energy can be written as $G=M-TS$. We have
\begin{equation}
\label{2-13}
G_{\rm H}=M_{\rm H}-T_{\rm H}S=\frac{(l^{2}+r_{\rm +}^{2})(q^{3}+r_{\rm +}^{3})}{2l^{2}r_{\rm +}^{2}}-\frac{r_{+}\Big(3r_{\rm +}^{5}+l^{2}(r_{\rm +}^{3}-2q^{3})\Big)}{4l^{2}(r_{\rm +}^{3}+q^{3})},
\end{equation}
\begin{eqnarray}
\label{2-14}
G_{\rm B}=M_{\rm B}-T_{\rm B}S=\frac{(l^{2}+r_{\rm +}^{2})(q^{2}+r_{\rm +}^{2})^{3/2}}{2l^{2}r_{\rm +}^{2}}-\frac{r_{\rm +}\Big(3r_{\rm +}^{4}+l^{2}(r_{\rm +}^{2}-2q^{2})\Big)}{4l^{2}(r_{\rm +}^{2}+q^{2})}.
\end{eqnarray}
In this section, we calculated the equation of the state, the heat capacity, and the Gibbs free energy of these two BHs. Next, we mainly analyze two thermodynamic phase transition characteristics.

\section{$P-\upsilon$ critical phase transition and the cooling-heating phase transition}
\label{sec3}
\subsection{$P-\upsilon$ critical phase transition}
\label{sec3-1}
The phase transition critical condition is ${\partial P}/{\partial \upsilon}=0={{\partial}^{2}P}/{\partial \upsilon^{2}}$ \cite{18}, where $\upsilon$ is the specific volume, satisfying $\upsilon=2r_{\rm +}$. Based on the critical condition and equation (\ref{2-8}), the critical volume $\upsilon_{\rm c}$, critical pressure $P_{\rm c}$, and critical temperature $T_{\rm c}$ of the HAdS BH are obtained, we have
\begin{eqnarray}
\label{3-1-1}
\upsilon_{\rm Hc}=2 (14+6\sqrt{6})^{1/3}g,\\
P_{\rm Hc}=\frac{3(3+\sqrt{6})}{2^{14/3}(7+3\sqrt{6})^{5/3} \pi g^{2}},\\
T_{\rm Hc}=\frac{(5\sqrt{2}-4\sqrt{3})(7+3\sqrt{6})^{2/3}}{2^{17/6} \pi g}.
\end{eqnarray}
The universal constant of the HAdS BH as
\begin{equation}
\label{3-1-2}
\varepsilon_{\rm H}=\frac{P_{\rm Hc}\upsilon_{\rm Hc}}{T_{\rm Hc}}=0.393031.
\end{equation}
\par
For the BAdS BH, the critical thermodynamic quantity are obtained according to equation (\ref{2-9}), one can get
\begin{eqnarray}
\label{3-1-3}
\upsilon_{\rm Bc}=\sqrt{2(15+\sqrt{273})}g,\\
P_{\rm Bc}=\frac{27+\sqrt{273}}{12(15+\sqrt{273})^{2}\pi g^{2}},\\
T_{\rm Bc}=\frac{(17-\sqrt{273})\sqrt{\frac{1}{2}(15+\sqrt{273})}}{24\pi g}.
\end{eqnarray}
The BAdS BH universal constant as
\begin{equation}
\label{3-1-4}
\varepsilon_{\rm B}=\frac{P_{\rm Bc}\upsilon_{\rm Bc}}{T_{\rm Bc}}=0.367069.
\end{equation}
These results are similar to the vdW system and the Reissner-Nordstr\"{o}m (RN)-AdS BH ($\varepsilon_{\rm vdW}=\varepsilon_{\rm R} = 0.375$) \cite{18}. The ratio of the HAdS BH is larger than 0.375, but the BAdS BH is smaller it. The $\varepsilon$ and $\upsilon_{\rm c}$ show a positive proportional relationship at a fixed critical temperature and pressure. Taking the RN-AdS BH as the standard, the HAdS BH is larger BH, and the BAdS BH is smaller BH. The larger BH shrinks towards the RN-AdS BH at the critical point, but the smaller BH expands to the RN-AdS BH. Therefore, we believe that the phase transition direction of two regular BHs is opposite. Note that the larger/smaller BHs here do not represent the large/small phase transition of the RN-AdS BH, which is only relative to the RN-AdS BH critical radius under certain conditions.

\par
The $P-\upsilon$ isotherm curves using equations (\ref{2-8}) and (\ref{2-9}) are shown in Figure 1. Two-phase transition branches exist below the critical temperature ($T<T_{\rm c}$), one is in the small radius region (corresponding to fluid phase in the vdW system), and the other is in the large radius region (corresponding to the gas phase). The critical isotherm ($T=T_{\rm c}$) has an inflection point, corresponding to the critical pressure. Above the critical isotherm ($T>T_{\rm c}$), there is no inflection point, and pressure is a monotonically decreasing function of radius. This phase transition is the first order for $T<T_{\rm c}$, while it becomes second order at $T_{\rm c}$ just as the same as the case in the vdW system.
\begin{center}
\includegraphics[width=9cm,height=6cm]{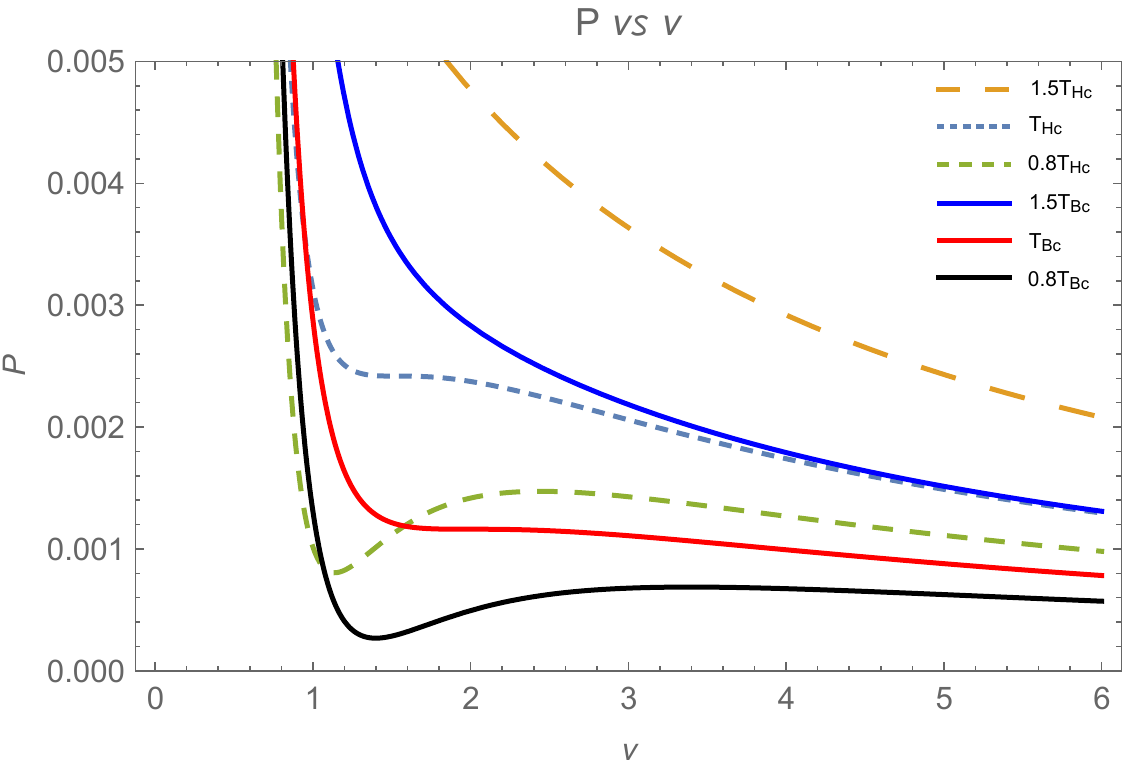}
\parbox[c]{15.0cm}{\footnotesize{\bf Fig~1.}
$P-\upsilon$ diagram for the HAdS BH and the BAdS BH with $g=1$. The dotted curves correspond to the HAdS BH, the solid curves represent the BAdS BH.}
\label{fig1}
\end{center}

\par
Taking the vdW phase transition as standard, $\omega_{\rm H(B)} \equiv (\varepsilon_{\rm H(B)}-\varepsilon_{\rm vdW})/\varepsilon_{\rm vdW}$ represent the BH system produces a deviation similar to the vdW gas-liquid phase transition, providing the criterion of the various BHs system to produce the vdW like system phase transition \cite{Cai}. Take the RN-AdS BH as an example, $\omega_{\rm R}=0$, it implies that the phase transition of the RN-AdS BH is entirely similar to that of the vdW system. For our BHs, the HAdS BH is $\omega_{\rm H}=0.048$ and the BAdS BH is $\omega_{\rm B}=0.021$, which means that the phase transition of the BAdS BH tends to be more the van der Waals (vdW) phase transition.

\par
Considering the innermost stable circular orbits of these two regular BHs, $r_{\rm isco}$ represents the minimum radius of particle rotation around BH. The BH phase transition always takes place on the side of the innermost stable circular orbit and finally tends to be stable. The calculation formula of the innermost stable circular orbit as
\begin{equation}
\label{3-1-5}
r_{\rm isco}=\frac{3f(r_{\rm isco})f'(r_{\rm isco})}{2f'(r_{\rm isco})^{2}-f(r_{\rm isco})f''(r_{\rm isco})}.
\end{equation}
Assuming $(M,~l,~g)$ is $(1,~1,~0.4)$, the innermost stable circular orbits of the HAdS BH is $r_{\rm Hisco}\simeq 3.74587$, the BAdS BH is $r_{\rm Bisco}\simeq 3.50254$, and the RN-AdS BH is $r_{\rm Risco}\simeq 3.62808$. Taking the RN-AdS BH as the standard, the HAdS and BAdS BHs have opposite directions. This results are self-consistent with our analysis above.

\subsection{Cooling-heating phase transition}
\label{sec3-2}
\par
The JT expansion is an isenthalpic process in the extended phase space. The JT coefficient $\mu \equiv {({\partial T}/{\partial P})}_{\rm H}$ determined cooling-heating regions \cite{25}. It is can be written as
\begin{equation}
\label{3-2-1}
\mu={\Big(\frac{\partial T}{\partial P}\Big)}_{\rm H}=\frac{1}{C_{\rm P}}\Big[T \Big({\frac{\partial V}{\partial T}}\Big)_{\rm P}-V\Big].
\end{equation}
Setting $\mu=0$, the inversion temperature is obtained, i.e.,
\begin{equation}
\label{3-2-2}
T_{\rm i}=V {\Big(\frac{\partial T}{\partial V}\Big)}_{\rm P}.
\end{equation}

\par
According to equations (\ref{2-4}) and (\ref{2-5}), the pressure $P$ and temperature $T$ can be reorganized as a function of $M$ and $r_{\rm +}$, we have
\begin{equation}
\label{3-2-3}
P_{\rm H}(M,{r}_{\rm +})=\frac{3(2M r_{\rm +}^{2}-g^{3}-r_{\rm +}^{3})}{8 \pi r_{\rm +}^{2}(g^{3}+r_{\rm +}^{3})},
\end{equation}
\begin{equation}
\label{3-2-4}
T_{\rm H}(M,{r}_{\rm +})=\frac{3 M r_{\rm +}^{5}-r_{\rm +}^{6}-2g^{3}r_{\rm +}^{3}-g^{6}}{2 \pi r_{\rm +}(g^{3}+r_{\rm +}^{3})^{2}},
\end{equation}
and the JT coefficient of the HAdS BH can be expressed as
\begin{eqnarray}
\label{3-2-5}
\mu_{\rm H}&&={\Big(\frac{\partial T}{\partial r_{\rm +}}\Big)}_{\rm M} / {\Big(\frac{\partial P}{\partial r_{+}}\Big)}_{\rm M}\nonumber\\
&&=\frac{4r_{\rm +}\Big[g^{6}+8g^{3}(r_{\rm +}^{3}+2P_{\rm i} \pi r_{\rm +}^{5})-2(r_{\rm +}^{6}+4P_{\rm i} \pi r_{\rm +}^{8})\Big]}{3(g^{3}+r_{\rm +}^{3})(2g^{3}-r_{+}^{3}-8P_{\rm i} \pi r_{\rm +}^{5})}.
\end{eqnarray}
Setting $\mu=0$, one can get the positive and real root which has a physical meaning, and the corresponding pressure in the root is the inversion pressure $P_{\rm i}$. Substituting the root into equation (\ref{2-5}), we have the inversion temperature of the HAdS BH (since these expressions are too complex, they are not explicitly given here). For the BAdS BH, we obtained the inversion temperature and plotted the invertion curves in Ref. \cite{28}. From Figure 2, we can see that the inversion curves of the HAdS BH are always higher than that of the BAdS BH under the same pressure and magnetic charge. In comparison with vdW liquid system, the regular BHs show a one inversion curve and not closed. One can observe that the increase of g value leads to the inversion temperature of BH to increase in the given pressure.
\begin{center}
\includegraphics[width=9cm,height=6cm]{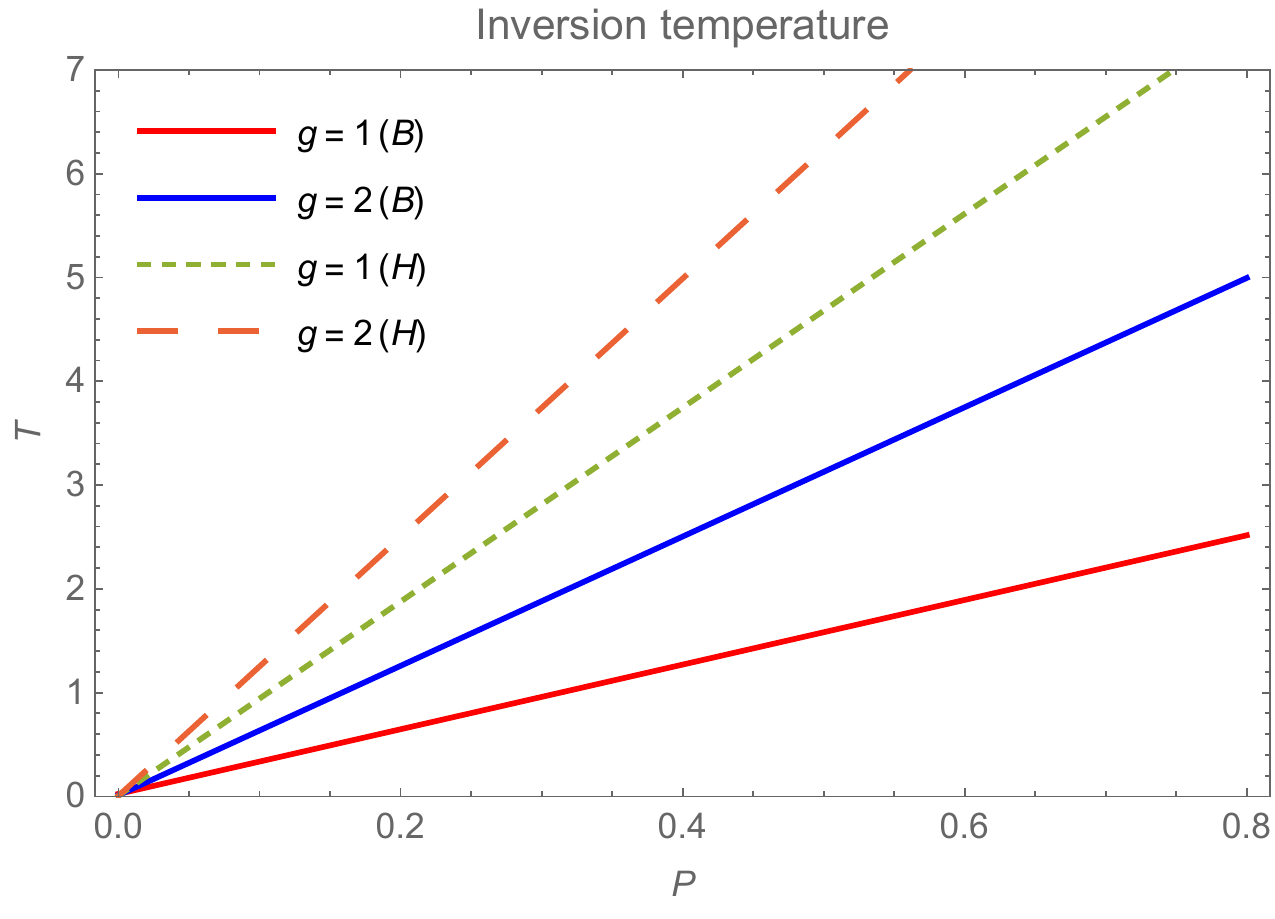}
\parbox[c]{15.0cm}{\footnotesize{\bf Fig~2.}  
The inversion curves of the regular BHs. The dotted curves correspond to the HAdS BH; the red and blue curves correspond to the BAdS BH.}
\label{fig2}
\end{center}

\par
The isenthalpic and inversion curves of the HAdS BH and the BAdS BH are shown in Figure 3. It is found that the isenthalpic curves of the HAdS BH are higher than those of the BAdS BH. The intersection points of inversion and isenthalpic curves of the HAdS BH is higher than BAdS BH, which means that the HAdS BH needs a higher temperature to produce cooling-heating phase transition on the same pressure.
\begin{center}
\includegraphics[width=9cm,height=6cm]{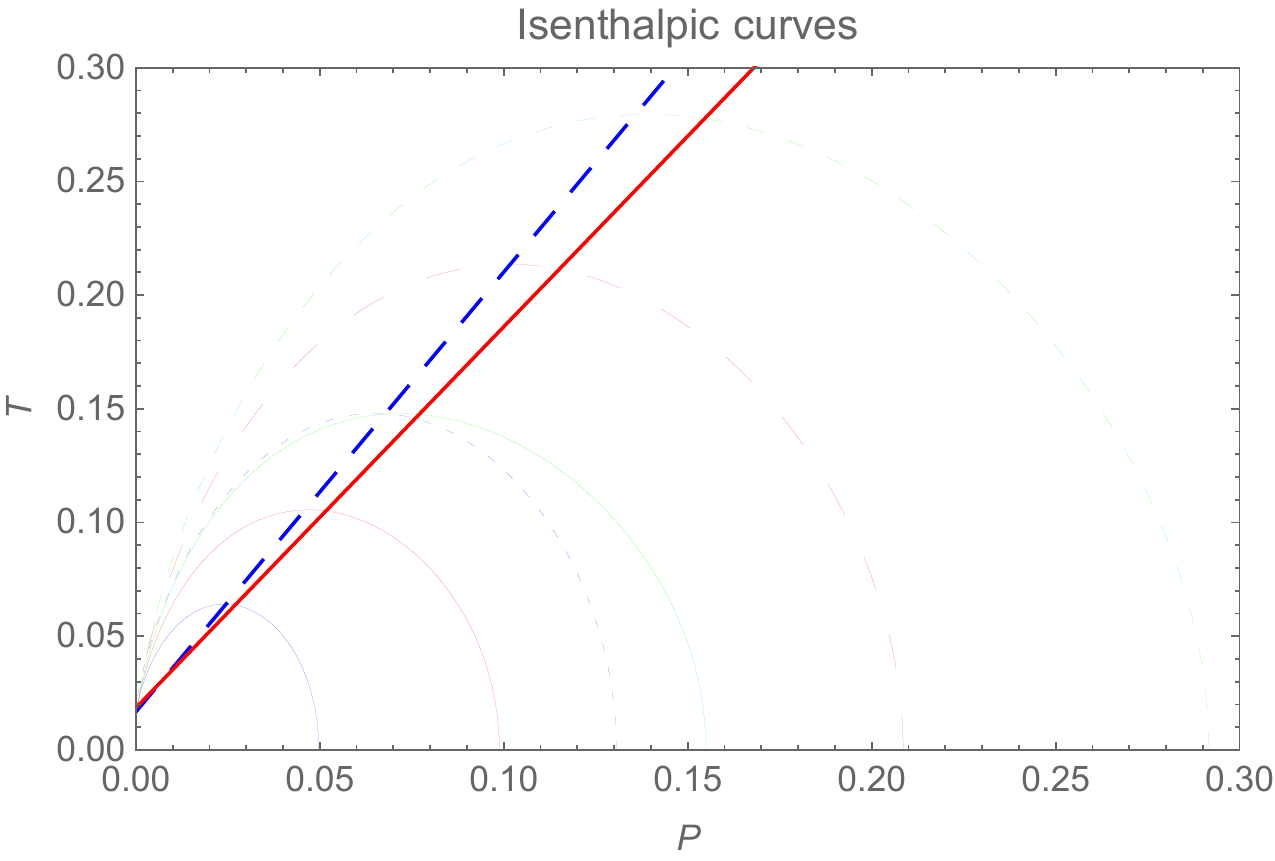}
\parbox[c]{15.0cm}{\footnotesize{\bf Fig~3.}  
The isenthalpic and inversion curves when $g=1$. The dotted curves correspond to the isenthalpic curves for the HAdS BH with $M=2,2.5,3$. The solid curves correspond to the isenthalpic curves for the BAdS BH with same mass. Correspondingly, the blue dotted straight line and red solid line are inversion curves.}
\label{fig3}
\end{center}
The minimum inversion temperature can be obtained by demanding $P_{\rm i}=0$. Based on equations (\ref{2-5}) and (\ref{3-2-5}), we obtain the minimum inversion temperature of the HAdS BH,
\begin{equation}
\label{3-2-6}
T_{\rm Hi}^{\rm min}=\frac{1}{2^{7/6}(4+3\sqrt{2})^{1/3}(2+\sqrt{2})\pi q}.
\end{equation}
The ratio between the minimum inversion temperature and the critical temperature of the HAdS BH can be calculated, i.e.,
\begin{equation}
\label{3-2-7}
\frac{T_{\rm Hi}^{\rm min}}{T_{\rm Hc}}=0.545685.
\end{equation}
This ratio of the BAdS BH is $0.536622$ \cite{28}. We found that the ratio of the BAdS BH is lower than the HAdS BH. They are both higher than those of singular BHs for this ratio, which depended on the space-time structure of BH. The regular BH exists only in the case of a repulsive de-Sitter core at the origin, and the large ratio for the regular-AdS BH may be due to the repulsive de-Sitter core near the origin of the regular BH.

\section{Some other comparisons}
\label{sec4}
\subsection{Smallest existence mass}
\label{sec4-1}
\par
Considering the magnetic charge as a minimum length of BH \cite{40,41}, it can be interpreted alternatively as a minimal cut-off length that makes gravity ultraviolet self-complete. We can re-written the equations (\ref{2-2}) and (\ref{2-3}) as
\begin{equation}
\label{4-1-1}
f_{\rm H}(x)=1+x^{2}-\frac{2 m x^{2}}{x^{3}+Q^{3}},
\end{equation}
\begin{equation}
\label{4-1-2}
f_{\rm B}(x)=1+x^{2}-\frac{2m x^{2}}{(x^{2}+Q^{2})^{\frac{3}{2}}},
\end{equation}
where $x\equiv{r}/{l}$, $m\equiv{M}/{l}$, $Q\equiv{g}/{l}$ \cite{40,41}. $f(x)$ as a function of $x$ are shown in Figure 4. It is observed that the BH cases will generically have an inner ($x_{\rm -}$) and outer ($x_{\rm +}$) Killing horizon, the two cases separated by an extreme BH with degenerate Killing horizon. Elementary analysis of the zero of $f(x)$ reveals a critical mass $m_{\rm 0}$ and radius $x_{\rm 0}$ such that, for $x>0$, $f(x)$ has no zeros if $m<m_{\rm 0}$, one double zero at $x=x_{0}$ if $m=m_{0}$, and two simple zeros at $x=x_{\pm}$ if $m>m_{0}$. A regular space-time with the same causal structure as flat space-time, a regular extreme BH with degenrate Killing horizon, and a regular non-extreme BH with both outer and inner Killing horizons. Regular BH can not from with mass $m<m_{0}$. Also, the inner horizon has radius $x_{-}>l$ which is very close to $l$ for all but the smallest existence masses. In this sense, the HAdS BH and the BAdS BH cores has a universal structure.

\par
According to the condition of the smallest existence mass
\begin{equation}
\label{4-1-3}
f(x)=\frac{\partial f(x)}{\partial x}=0,
\end{equation}
the HAdS BH smallest existence mass is $m_{\rm H0}=0.459$ and the BAdS BH smallest existence mass is $m_{\rm B0}=0.648$ for the $Q=0.4$. The smallest existence mass of the HAdS BH is smaller than that of the BAdS BH, meaning that the inner horizon and the outer horizon of his BH are easier to merge, and the singularity is easier to expose.
\begin{center}
\includegraphics[width=9cm,height=6cm]{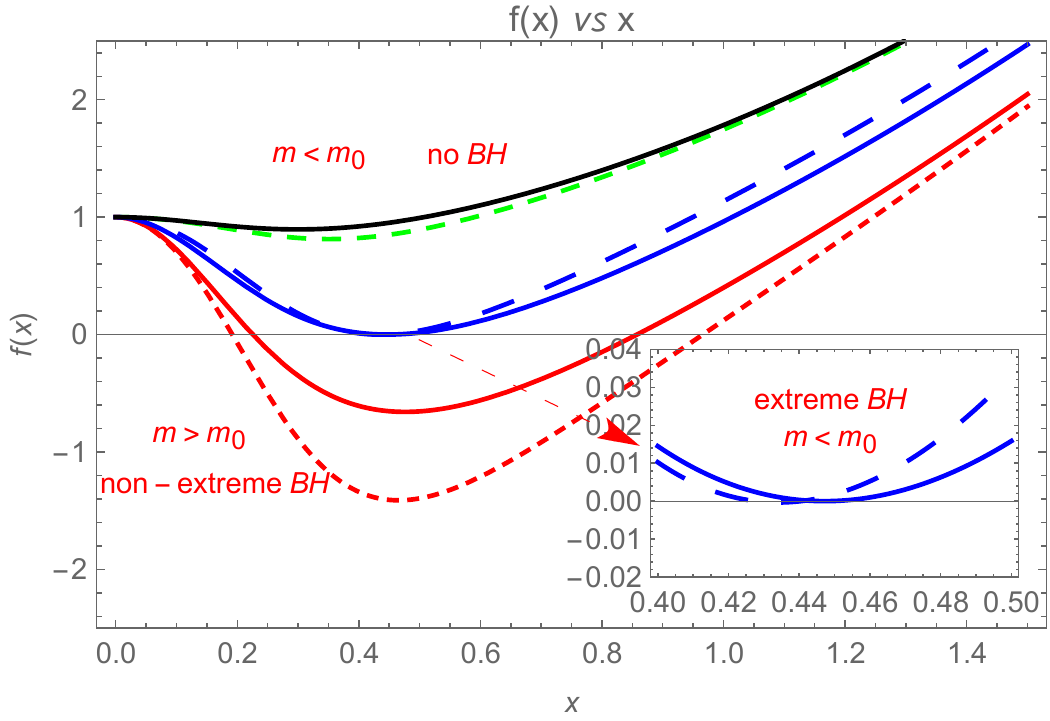}
\parbox[c]{15.0cm}{\footnotesize{\bf Fig~4.}  
$f(x)$ as a function of $x$ for $Q=0.4$. The dotted curves represents the HAdS BH and solid curves represents the BAdS BH. From top to bottom, it corresponds to $m=0.135$, $m=m_{0}$ and $m=1$ respectively.}
\label{fig4}
\end{center}

\subsection{Critical magnetic charge and zero-temperature remnants}
\label{sec4-2}
The temperature of the HAdS BH and the BAdS BH can be re-expressing according to equations (\ref{2-5}) and (\ref{2-7}), we have
\begin{equation}
\label{4-2-1}
\mathcal{T}_{\rm H}=\frac{3x_{\rm +}^{5}+x_{\rm +}^{3}-2Q^{3}}{4 \pi x_{\rm +}(Q^{3}+x_{\rm +}^{3})},
\end{equation}
and
\begin{equation}
\label{4-2-2}
\mathcal{T}_{\rm B}=\frac{3x_{\rm +}^{4}+x_{\rm +}^{2}-2Q^{2}}{4 \pi x_{\rm +}(Q^{2}+x_{\rm +}^{2})}.
\end{equation}
Based on the inspiration of critical behaviour $(\partial \mathcal{T} / \partial x_{\rm +})=0$, $(\partial^{2} \mathcal{T} / \partial x_{\rm +}^{2})=0$, the critical magnetic charge and zero-temperature remnant of these two regular BHs are obtained, respectively. The HAdS BH critical magnetic charge is $Q_{\rm Hc}=0.164755$, and the BAdS BH critical magnetic charge is $Q_{\rm Bc}=0.131736$. The temperature $\mathcal{T}$ as a function of $x_{+}$ are presented in Figure 5. We can see that the first order phase transition take place between small/large stable BHs when the values below the critical magnetic charge ($Q<Q_{\rm c}$). The temperature is a monotonically increasing function of radius for $Q>Q_{\rm c}$. The zero-temperature remnants of the BAdS BH are always smaller than that of the HAdS BH when the $Q$ is a constant.
\begin{center}
\includegraphics[width=9cm,height=6cm]{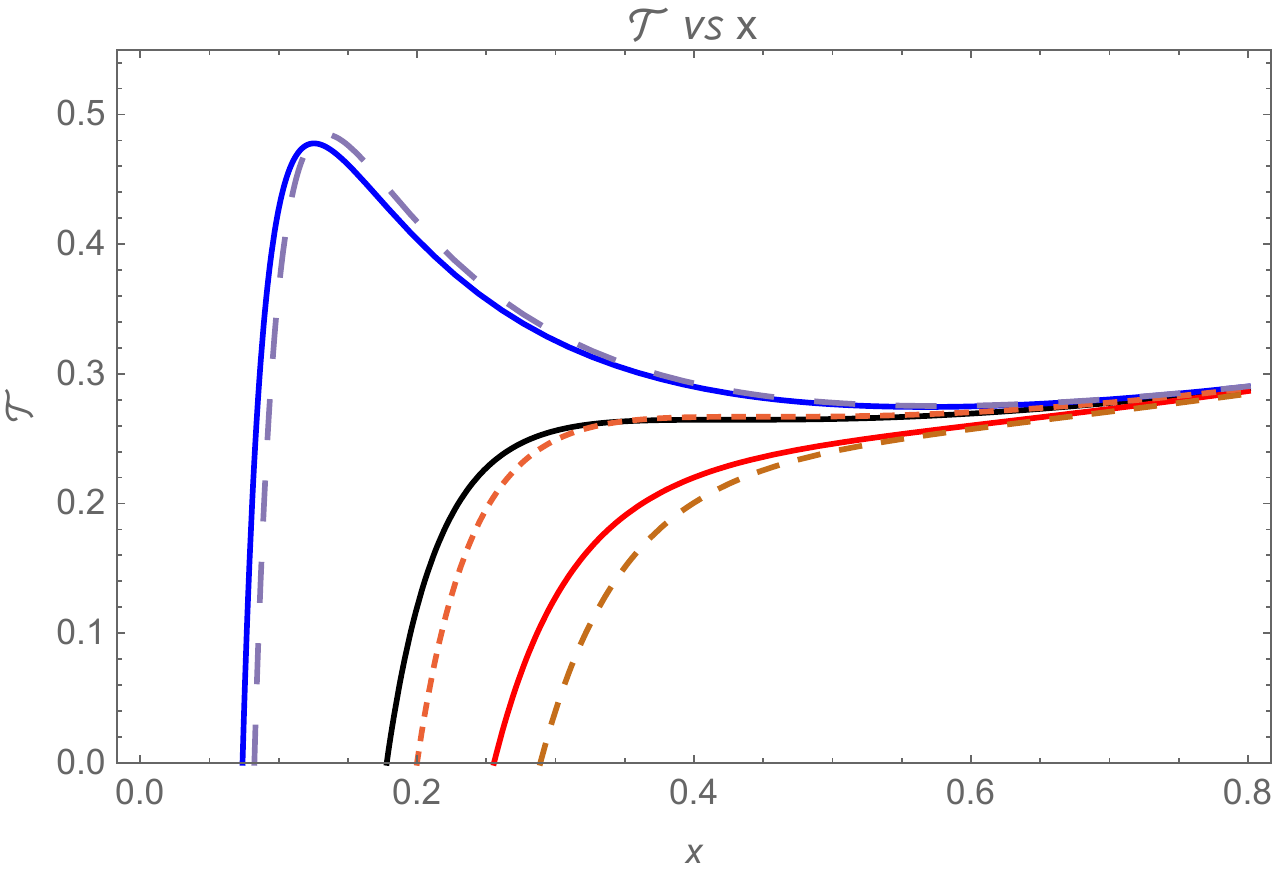}
\parbox[c]{15.0cm}{\footnotesize{\bf Fig~5.}  
The function $\mathcal{T}$ vs $x$. The dotted curves represent the HAdS BH and solid curves represent the BAdS BH corresponding $Q=0.4 Q_{\rm c}$, $Q=Q_{\rm c}$ and $Q=1.5 Q_{\rm c}$. The HAdS BH zero temperature remnant at $x=0.0826$, $x=0.2012$ and $x=0.2899$ and the BAdS BH zero temperature remnant at $x=0.0747$, $x=0.1771$ and $x=0.2573$.}
\label{fig5}
\end{center}

\section{Conclusion}
\label{sec:4}
\par
The thermodynamic behaviors of the regular HAdS BH and BAdS BH have been revealed in this analysis. By investigating the $P-\upsilon$ critical phase transition and cooling-heating phase transition, we found that the phase transition of the HAdS BH and BAdS BH is similar to that of the vdW system in the extended phase space. For the $P-\upsilon$ critical phase transition, the universal constant $\varepsilon_{H}$ of the HAdS BH is more significant than that of the vdW system, and the universal constant of the BAdS BH is small than that of the vdW system. Taking the RN-AdS BH as the standard, the phase transition direction of two regular BHs is opposite, and the phase transition of the BAdS BH tends to be more the van der Waals (vdW) phase transition.

\par
We investigated the cooling-heating phase transition of the HAdS BH and compared it with the BAdS BH. The inversion curves of the HAdS BH are always higher than that of the BAdS BH under the same pressure and magnetic charge. The result shows that the HAdS BH needs a higher temperature to produce cooling-heating phase transition at the same pressure. The ratio between the minimum inversion temperature and the critical temperature of these regular BHs is calculated. It is found that the ratio of the BAdS BH is lower than the HAdS BH. However, they are both higher than those of singular BHs for this ratio, which depends on the space-time structure of BH. The regular BH exists only in the case of a repulsive de-Sitter core at the origin, and the large ratio for the regular-AdS BH may be due to the repulsive de-Sitter core near the origin of the regular BH.

\par
For the smallest existence mass, we obtained that the smallest existence mass of the HAdS BH is smaller than that of the BAdS BH, which means that the inner and outer horizons of his BH are easier to merge and the singularity is easier to expose. We found that the BAdS BH is always smaller than that of the HAdS BH for the critical magnetic charge and zero-temperature remnants.

\section*{Acknowledgments}
The authors would like to thank the anonymous reviewers for their helpful comments and suggestions, which helped to improve the quality of this paper. This work is supported by the National Natural Science Foundation of China (Grant No.12133003, 11851304, and U1731239), by the Guangxi Science Foundation and special funding for Guangxi distinguished professors (2017AD22006).

\clearpage

\end{CJK}

\section{References}
\addcontentsline{toc}{chapter}{References}
\begin{thebibliography}{99}\footnotesize
\itemsep=-3pt plus.2pt minus.2pt   

\bibitem{1}
S. W. Hawking and G. F. R. Ellis, (1973) Cambridge.

\bibitem{2}
J. M. Bardeen, Proceedings of GR5, Tbilisi, Georgia, USSR, (1968) p. 174.

\bibitem{3}
E. Ay\'{o}n-Beato and A. Garc\'{\i}a, Phys. Rev. Lett {\bf 80} (1998) 5056; Gen. Rel. Grav. {\bf 31} (1999) 629; Phys. Lett. B. {\bf 25} (1999) 464; Phys. Lett. B. {\bf 493} (2000) 149.

\bibitem{4}
S. A. Hayward, Phys. Rev. Lett. {\bf 96} (2006) 031103.

\bibitem{5}
M. Azreg-A\"{\i}nou, Phys. Rev. D, {\bf 90} (2014) 064041.

\bibitem{6}
E. L. B. Junior, M. E. Rodrigues and M. J. S. Houndjo, JCAP, {\bf 1510} (2015) 060.

\bibitem{7}
A. Flachi, J. P. S. Lemos, Phys. Rev. D, {\bf 87} (2013) 024034.

\bibitem{8}
M. S. Ma, R. Zhao, Class. Quant. Grav, {\bf 31} (2014) 245014.

\bibitem{9}
J. Y. Man, H. B. Cheng, Gen. Rel. Grav {\bf 46} (2014) 1559.

\bibitem{10}
Z. Y. Fan, Eur. Phys. J. C, {\bf 77} (2017) 266.

\bibitem{11}
S. H. Mehdipour, M. H. Ahmadi, Nucl. Phys. B, {\bf 49} (2018) 926.

\bibitem{12}
J. D. Bekenstein. Lett. Nuovo Cimento, {\bf 4} (1972) 737; Phys. Rev. D, {\bf 7} (1973) 2333; Phys. Rev. D, {\bf 74} (1974) 3292.

\bibitem{13}
J. M. Bardeen, B. Carter, S. W. Hawking, Commun. Math. Phys {\bf 31} (1973) 161.

\bibitem{14}
S. W. Hawking, Nature, 30 (1974) 248; Commun. Math. Phys, {\bf 43} (1975) 199.

\bibitem{15}
S. W. Hawking, D. N. Page, Commun. Math. Phys, {\bf 87} (1983) 577.

\bibitem{16}
A. Chamblin, R. Emparan, C. V. Johnson, R. C. Myers, Phys. Rev. D, {\bf 60} (1999) 064018; Phys. Rev. D, {\bf 60} (1999) 104026.

\bibitem{17}
B. P. Dolan, Class. Quant. Grav, {\bf 28} (2011) 235017.

\bibitem{18}
S. Gunasekaran, R. B. Mann, D. Kubiznak. JHEP, {\bf 1211}: 110 (2012).

\bibitem{19}
R. G. Cai, L. M. Cao, L. Li, R. Q. Yang, JHEP, {\bf 09} (2013) 005.

\bibitem{20}
A. M. Frassino, D. Kubiznak, R. B. Mann, F. Simovic, JHEP, {\bf 09} (2014) 80.

\bibitem{21}
J. Xu, L. M. Cao, Y. P. Hu, Phys. Rev. D, {\bf 91} (2015) 124033.

\bibitem{22}
S. W. Wei, P. Cheng, Y. X. Liu, Phy. Rev. D, {\bf 93} (2016) 084015.

\bibitem{23}
G. Miao, Z. M. Xu, Phys. Rev. D, {\bf 8} (2018) 084051.

\bibitem{24}
R. Li, J. Wang, Phys. Lett. B, {\bf 813} (2021) 136035.

\bibitem{25}
\"{O}. \"{O}kc\"{u}, E. Ayd{\i}ner, Eur. Phys. J. C, {\bf 77} (2017) 24; Eur. Phys. J. C, {\bf 78} (2018) 123.

\bibitem{26}
J. X. Mo , G. Q. Li, S. Q. Lan, X. B. Xu, Phys. Rev. D, {\bf 98} (2018) 124032.

\bibitem{27}
S. Q. Lan, Phys. Rev. D, {\bf 98} (2018) 084014.

\bibitem{28}
S. Guo, J. Pu, Q. Q. Jiang, X. T. Zu, Chin. Phys. C, {\bf 44} (2020) 035102.

\bibitem{29}
S. Guo, Y. Han, G. P. Li. Class. Quant. Grav, {\bf 37} (2020) 042001; Mod. Phys. Lett. A, {\bf 35} (2020) 2050113.

\bibitem{30}
X. X. Zeng, Y. W. Han, D. Y. Chen, Chin. Phys. C, {\bf 10} (2019) 105104.

\bibitem{31}
X. X. Zeng, X. Y. Hu, K. J. He, Nucl. Phys. B, {\bf 949} (2019) 114823.

\bibitem{32}
K. J. He, X. Y. Hu, X. X. Zeng, Chin. Phys. C, {\bf 43} (2019) 125101.

\bibitem{33}
C. V. Johnson, Class. Quant. Grav, {\bf 31} (2014) 205002.

\bibitem{34}
A. Chakraborty, C. V. Johnson, Int. J. Mod. Phys. D, {\bf 27} (2018) 1950012.

\bibitem{35}
S. Guo, Y. L. Huang, K. J. He, G. P. Li, Mod. Phys. Lett. A, {\bf 36} (2021) 2150108.

\bibitem{L}
S. Guo, E. W. Liang, Class. Quant. Grav, {\bf 38} (2021) 125001.

\bibitem{36}
E. Ay\'{o}n-Beato and A. Garc\'{\i}a. Phys. Rev. Lett, {\bf 80} (1998) 5056; Gen. Rel. Grav, {\bf 31} (1999) 629; Phys. Lett. B, {\bf 464} (1999) 25; Phys. Lett. B, {\bf 493} (2000) 149.

\bibitem{37}
M. Cvetic, G. W. Gibbons, D. Kubiznak and C. N. Pope. Phys. Rev. D, {\bf 84} (2011) 024037.

\bibitem{38}
Z. Y. Fan, X. B. Wang, Phys. Rev. D, {\bf 94} (2016) 124027.

\bibitem{39}
S. H. Mehdipour, M. H. Ahmadi, Nucl. Phys. B, {\bf 49} (2018) 926.

\bibitem{Cai}
R. G. Cai, L. M. Cao, et al., JHEP, {\bf 09} (2013) 005.

\bibitem{40}
A. G. Tzikas. Phys. Lett. B, {\bf 788} (2019) 219.

\bibitem{41}
C. L. Ahmed. Rizwan, Naveena,at.al. arXiv:1811.10838.


\end{thebibliography}
\end{document}